\documentclass[conference]{IEEEtran}
\IEEEoverridecommandlockouts
\usepackage{amsmath,amssymb,amsfonts}
\usepackage{algorithmic}
\usepackage{graphicx}
\usepackage{textcomp}
\usepackage{xcolor}
\def\BibTeX{{\rm B\kern-.05em{\sc i\kern-.025em b}\kern-.08em
    T\kern-.1667em\lower.7ex\hbox{E}\kern-.125emX}}
\begin{document}
  
\title{Can LLMs Be Trusted for Evaluating RAG Systems? A Survey of Methods and Datasets}

\author{\IEEEauthorblockN{1\textsuperscript{st} Lorenz Brehme}
\IEEEauthorblockA{\textit{Department of Computer Science} \\
\textit{University of Innsbruck}\\
Innsbruck, Austria \\
https://orcid.org/0009-0009-4711-2564}
\and
\IEEEauthorblockN{2\textsuperscript{nd} Thomas Ströhle}
\IEEEauthorblockA{\textit{Department of Computer Science} \\
\textit{University of Innsbruck}\\
Innsbruck, Austria \\
https://orcid.org/0000-0002-1954-6412}
\and
\IEEEauthorblockN{3\textsuperscript{rd} Ruth Breu}
\IEEEauthorblockA{\textit{Department of Computer Science} \\
\textit{University of Innsbruck}\\
Innsbruck, Austria \\
https://orcid.org/0000-0001-7093-4341}

}
\IEEEpubid{\makebox[\columnwidth]{This paper has been accepted for presentation at the SDS25.\,\copyright~2025 IEEE.\hfill}%
\hspace{\columnsep}\makebox[\columnwidth]{ }}

\maketitle
\vspace{-2ex}

\begin{abstract}
Retrieval-Augmented Generation (RAG) has advanced significantly in recent years. The complexity of RAG systems, which involve multiple components—such as indexing, retrieval, and generation—along with numerous other parameters, poses substantial challenges for systematic evaluation and quality enhancement. Previous research highlights that evaluating RAG systems is essential for documenting advancements, comparing configurations, and identifying effective approaches for domain-specific applications. This study systematically reviews 63 academic articles to provide a comprehensive overview of state-of-the-art RAG evaluation methodologies, focusing on four key areas: datasets, retrievers, indexing and databases, and the generator component. We observe the feasibility of an automated evaluation approach for each component of a RAG system, leveraging an LLM capable of both generating evaluation datasets and conducting evaluations. In addition, we found that further practical research is essential to provide companies with clear guidance on the do's and don'ts of implementing and evaluating RAG systems. By synthesizing evaluation approaches for key RAG components and emphasizing the creation and adaptation of domain-specific datasets for benchmarking, we contribute to the advancement of systematic evaluation methods and the improvement of evaluation rigor for RAG systems. Furthermore, by examining the interplay between automated approaches leveraging LLMs and human judgment, we contribute to the ongoing discourse on balancing automation and human input, clarifying their respective contributions, limitations, and challenges in achieving robust and reliable evaluations.
\end{abstract}

\section{Introduction}
In recent years, Large Language Models (LLMs) have made significant progress in research and have grown increasingly popular \cite{gao_retrieval-augmented_2024}. However, LLMs face several challenges, including issues with hallucinations caused by insufficient context \cite{zhang_sirens_2023}, as well as limitations in their learned content, which prevent them from addressing questions requiring specific or proprietary information \cite{gao_retrieval-augmented_2024}. To address these issues, \cite{lewis_retrieval-augmented_2021} introduced Retrieval-Augmented Generation (RAG), which extends LLMs by integrating external knowledge sources for knowledge-intensive natural language processing (NLP) tasks. By incorporating domain-specific information, RAG systems enable tailored responses for specialized topics, improving accuracy, relevance, and contextual understanding. Since 2022, numerous RAG systems have demonstrated the effectiveness of this approach in overcoming some limitations of LLMs \cite{gao_retrieval-augmented_2024}. These systems are applied across a wide range of NLP tasks and outperform in domain-specific scenarios by leveraging specialized knowledge to enhance performance and relevance \cite{gao_retrieval-augmented_2024}.
RAG systems operate through three interconnected components: (1) indexing, which structures and organizes external knowledge bases; (2) retrieval, which identifies and extracts relevant documents from these sources; (3) and generation, which combines retrieved information with the input to produce a coherent, contextually relevant response using an LLM and prompt engineering techniques. These systems offer numerous customization options, including fine-tuning retrieval mechanisms, optimizing prompting techniques, refining generation models, and customizing knowledge base design \cite{lyu_crud-rag_2024, kukreja_performance_2024}. This flexibility raises the question of what settings should be used to configure RAG systems, and how these systems can be compared and evaluated to determine the most effective system for specific domains.
This particular task of identifying optimal parameters in RAG systems is commonly referred to in the literature as \textit{RAG evaluation} \cite{es_ragas_2023, saad-falcon_ares_2024}. During the RAG evaluation, each component of the RAG system is systematically evaluated within a predefined workflow to ensure that the system meets the overall quality standards \cite{es_ragas_2023}. The process begins with an input query from a prepared QA evaluation dataset, which is used to compute metrics such as retrieval accuracy and response relevance, providing statistical insight into the performance of the RAG components and thus a comprehensive understanding of their effectiveness \cite{es_ragas_2023}.
RAG evaluation presents several challenges: A major hurdle is the definition of robust methods for assessing the quality of the system's responses; frameworks such as RAGAS \cite{es_ragas_2023} and ARES \cite{saad-falcon_ares_2024} provide comprehensive metrics for evaluating the generator's responses in terms of relevance, accuracy, and overall performance. Retriever evaluation focuses on evaluating the relevance of retrieved documents and determining whether the selected chunks effectively contribute to answering the query \cite{salemi_evaluating_2024}, while indexing evaluation primarily emphasizes performance metrics such as indexing and retrieval speed \cite{kukreja_performance_2024, caspari_beyond_2024}. Equally critical is the selection of an appropriate database that both contains domain-specific information and is complex enough to effectively test the system's ability to meet quality benchmarks \cite{tang_multihop-rag_2024, wang_domainrag_2024}. Furthermore, creating QA evaluation datasets requires significant human effort and domain expertise, so they are often augmented with an artificial dataset generated by an LLM using pieces of given domain knowledge to streamline the process \cite{ es_ragas_2023, wang_domainrag_2024, pu_customized_2024}.
Previous literature reviews have either focused exclusively on evaluation of the retriever and generator components, overlooking indexing or broader aspects of dataset generation and enhancement \cite{yu_evaluation_2024}, or focused primarily on evaluation metrics and existing datasets, similarly omitting evaluation of indexing \cite{knollmeyer_benchmarking_2024}. However, there is a growing body of research emphasizing the need for dataset generation and enhancement \cite{tang_multihop-rag_2024}, the importance of including indexing strategies and their impact on system performance \cite{caspari_beyond_2024}, and the importance of automation through LLMs \cite{es_ragas_2023} in RAG evaluation.
The aim of this paper is therefore to conduct a systematic literature review (SLR) on the evaluation of RAG systems, synthesizing existing research to identify best practice and provide guidance on effective evaluation methodologies. We focus on examining datasets and the processes used to produce and improve them, with the aim of understanding their impact on the reliability of evaluations. Our analysis examines the evaluation of indexing, retrieval and generation components, highlighting the metrics and methodologies used. We also assess the interplay between automation, particularly through LLM, and human judgement, clarifying their roles and limitations in the evaluation process. By distinguishing between RAG tasks and their evaluation techniques, we offer insights into the unique challenges associated with each task type. Finally, we offer a coherent overview that integrates these findings and advances the understanding of effective RAG system evaluations.

In Section \ref{method}, we outline our approach to selecting relevant literature and conducting the SLR. The subsequent sections examine the evaluation of each RAG component. We begin with the dataset evaluation in Section \ref{evaluationdatasets}, followed by an assessment of indexing and database performance in Section \ref{evalauationindexing}. Next, we evaluate the retriever and its methodologies in Section \ref{evaluationretriever}. Finally, in Section \ref{evalautionofgenerator}, we analyze the generator tasks within the system. We conclude with a summary of the current state of RAG evaluation.

\section{Reasearch Method}
\label{method}

For our analysis, we conducted an SLR following the guidelines of \cite{kitchenham_guidelines_2007}. The process began with identifying relevant studies using \textit{advanced search} tools in online databases. We selected keywords and formulated a search query based on two key frameworks: RAGAS, an early automated evaluation approach for RAGs \cite{es_ragas_2023}, and \cite{tang_multihop-rag_2024}, which benchmarks RAG systems for multi-hop questions. Boolean operators were applied, and the query was iteratively refined to improve relevance.
The final search query was: \textit{"RAG" OR "Retrieval-Augmented Generation" OR "Retrieval Augmented Generation" OR "retriev* augment* generation"; AND Evaluation OR "Quality Assessment" OR "Benchmark" OR "Performance Evaluation" OR "evaluat*"}. Searches were conducted on November 11, 2024, with papers restricted to those published from 2021 onward. The search covered ACM, IEEE, arXiv, Elsevier, Google Scholar, and Web of Science—databases commonly used in computer science research \cite{kitchenham_guidelines_2007}.
This initial search yielded 71 papers: 50 from arXiv, four from ACM, one from Elsevier, nine from Google Scholar, four from IEEE, two from Springer, and one from Web of Science. We reviewed abstracts and excluded papers that did not primarily focus on evaluating or benchmarking RAG systems, narrowing the selection to 48 papers. A forward and backward search using Google Scholar added 21 more papers, of which six did not meet our criteria. Ultimately, our SLR included 63 relevant papers.
Previous literature reviews analyzed only 12 papers, with \cite{yu_evaluation_2024} focused on retriever and generator evaluation but omitted indexing and dataset enhancement, while \cite{knollmeyer_benchmarking_2024} provided an overview of evaluation datasets and methods, primarily discussing metrics without addressing indexing evaluation.
To structure our findings, we categorized papers based on their focus: evaluation of retrievers, generators, or embeddings. Additionally, we grouped them by dataset usage, distinguishing between generated datasets, enhanced datasets, and pre-existing datasets.

\section{Evaluation Datasets}
\label{evaluationdatasets}

In this survey, we found 87 existing question and answer (QA) datasets that were used to benchmark RAG systems. These datasets differ in the types of questions, the domain—such as legal \cite{pipitone_legalbench-rag_2024}, medicine \cite{xiong_benchmarking_2024}, or energy \cite{meyur_weqa_2024}—the length of the questions, and the type of context they provide, each introducing unique challenges and requirements for RAG system evaluation. The datasets contain question-answer-context tuples, where the question serves as input to the RAG system, and the context and answer act as evaluation references. For example, for \textit{What is the capital of Switzerland?}, the context might be \textit{Bern is the capital of Switzerland}, and the answer \textit{Bern}. Questions are categorized to enable structured evaluation of the RAG system's capabilities.

We identified five distinct categories of questions as being used to evaluate RAG systems in this study: One prevalent type includes questions with \textbf{(1) short answers}, which typically consist of a single word, a number, or a Boolean variable. This category also encompasses single-choice and multiple-choice questions, often requiring straightforward responses \cite{oberst_how_2023}. In contrast, \textbf{(2) long answer questions} demand more detailed responses, providing logical and well-reasoned explanations \cite{ru_ragchecker_2024}. Building on the complexity of long-answer questions, \textbf{(3) multi-hop questions} extend the challenge by requiring the integration of several pieces of context to form a coherent answer. These questions test a RAG system’s capacity to utilize multiple contexts and engage in logical reasoning across several steps \cite{tang_multihop-rag_2024}. Some evaluations include \textbf{(4) error-type questions} designed to expose system weaknesses by presenting incorrect or illogical contexts, requiring the RAG system to identify inconsistencies and respond appropriately \cite{ravi_lynx_2024, saad-falcon_ares_2024}, or trigger specific error cases within the system \cite{xu_face4rag_2024, peng_rag-confusionqa_2024}. Datasets vary in structure, including questions, answers, context, and sometimes misleading information. \textbf{(5) Misleading context} datasets test a system’s ability to detect and handle false information.
In general, we observed that datasets for evaluating RAG systems require more complex questions and answers. To achieve this, multi-hop questions and long answers were utilized \cite{tang_multihop-rag_2024}. 

\label{datasetsforquestion}
There are existing datasets that are publicly available for QA tasks that are in different domains. The most commonly used datasets are HotPotQA, NaturalQuestions, and MSMarco, with HotPotQA being particularly prominent. It includes questions with long and short answers, multi-hop reasoning, and fake context, making it a highly complex dataset. These datasets are created based on publicly available knowledge, not proprietary knowledge, which  presents a challenge because the RAG system then becomes redundant since the LLM has already been trained on this knowledge \cite{kenneweg_retrieval_2024}. To avoid this issue, existing databases were modified, or a complete new dataset which could include the proprietary knowledge was created \cite{kenneweg_retrieval_2024}.
\label{Creation of Datasets}
These new datasets are particularly important for RAG systems, which depend on domain-specific knowledge and are customized to meet the needs of specialized areas. As a result, evaluating them with public QA datasets is usually impractical. In such cases, custom datasets must be created manually \cite{schimanski_climretrieve_2024, han_rag-qa_2024}.
\paragraph{Creation by Humans}
In \cite{schimanski_climretrieve_2024, li_benchmarking_2024, han_rag-qa_2024}, human annotators develop datasets, with \cite{schimanski_climretrieve_2024, li_benchmarking_2024} focusing on questions crafted by humans in a specific field and subsequently evaluated by domain experts. One approach begins with 16 Yes/No questions on a domain-specific topic, iteratively refining concepts and developing complex, context-enriched questions \cite{schimanski_climretrieve_2024}.
When a QA dataset lacks complex answers, a human annotator formulates detailed responses, which are then reviewed by experts for a more comprehensive evaluation \cite{han_rag-qa_2024}.

\paragraph{Enhancing existing datasets}
Three methods were identified to enhance existing datasets and develop new ones.
For instance, when datasets contain only questions and context without answers, an LLM can generate the answers based on the provided context \cite{xu_face4rag_2024}. The second method generates answers to reflect specific error types to test evaluators and assess their robustness and reliability \cite{xu_face4rag_2024, peng_rag-confusionqa_2024, ravi_lynx_2024}.
Another method involves rewriting questions to enhance their complexity and suitability for RAG evaluation \cite{xie_rag_2024, wang_feb4rag_2024}. However, a risk with existing datasets is that LLMs may recall answers without using the given context, which is mitigated by evolving and regenerating questions for novelty \cite{qi_long2rag_2024}.

\paragraph{Generation of Datasets using an LLM}
\label{GenerationOfDataset}
Dataset creation is time-consuming; to address this, evaluation frameworks have been developed that generate datasets using LLMs. A given context was used to generate a specific question related to that context \cite{es_ragas_2023, tang_multihop-rag_2024}. Additionally, the LLM generated an answer corresponding to the question and context, completing the tuple. The resulting questions were then filtered to refine the dataset and improve its quality \cite{es_ragas_2023, tang_multihop-rag_2024}.
The generation process uses several prompting strategies, such as chain-of-thought prompting, along with specific metrics to derive the best questions for RAG evaluation \cite{es_ragas_2023, liu_cofe-rag_2024, krishna_fact_2024}.
It has been demonstrated that LLMs can answer questions using prior knowledge, even in the absence of context \cite{kenneweg_retrieval_2024}. To prevent this, post-training articles were used for question generation, except in domain-specific areas where LLMs lack prior exposure.

To enhance evaluation datasets, \cite{tang_multihop-rag_2024} proposed generating multi-hop questions requiring reasoning across multiple contexts. For instance,  \cite{pu_customized_2024} used one to five randomly sampled chunks from a domain-specific topic to create multi-hop queries. The approach was enhanced by using an LLM to add keywords to each chunk and QA pair to address missing critical information \cite{zhu_rageval_2024, liu_cofe-rag_2024, qi_long2rag_2024}.
In addition, one method addresses the generation of domain-specific questions by evaluating different aspects of domain-specific tasks: These datasets include a multi-hop dataset, a dataset focused on table information, and a dataset designed to test the faithfulness of RAGs, where LLMs generate questions from documents requiring specific knowledge \cite{wang_domainrag_2024, meyur_weqa_2024}.
Another method is to generate a noisy dataset\cite{wang_domainrag_2024, meyur_weqa_2024, saad-falcon_ares_2024, sun_multimodal_2024, chen_benchmarking_2023, ming_faitheval_2024}, where irrelevant, misleading, or conflicting context is added to the Question-Answer dataset. This approach aims to evaluate the ability of evaluators to detect hallucinations, assess the faithfulness of the RAG, and test the performance of the retriever.

Once the questions and answers have been created, they need to be filtered and selected for inclusion in the final dataset to ensure quality and accuracy by an LLM or manually by humans. In \cite{meyur_weqa_2024,liu_cofe-rag_2024, krishna_fact_2024, zhu_rageval_2024}, domain experts evaluate the generated questions for relevance and clarity.
In contrast, an LLM is used to evaluate datasets by assessing query-context alignment, answerability, query type relevance, or independence, assigning a score for each criterion \cite{tang_multihop-rag_2024, es_ragas_2023}.

\section{Indexing and Database Evaluation}
\label{evalauationindexing}
One crucial component of a RAG system is the database used by the retriever. Key factors such as indexing, embeddings, and chunk size significantly impact the system's performance. Two studies specifically focused on evaluating these components \cite{caspari_beyond_2024, kukreja_performance_2024}. In \cite{kukreja_performance_2024} the authors evaluated database performance using four key metrics: upload time, indexing time, retrieval speed, and throughput, providing quantitative performance measurements.
Additionally, \cite{yu_knowledge-centric_2024} evaluates different embeddings, chunking strategies and databases by analyzing their impact on overall performance.
In \cite{caspari_beyond_2024} they focus on evaluating embedding models, particularly in the context of RAG. The authors investigate the similarity of embedding models by comparing their representations and the similarity of retrieved results for specific queries. To measure the similarity of embeddings, \cite{caspari_beyond_2024} employs Centered Kernel Alignment (CKA). For evaluating the similarity of retrieved contexts, the Jaccard similarity coefficient was utilized and the RankSimilarity score to account the ranking of retrieved text chunks were introduced.
In total, the indexing component is primarily evaluated on performance metrics like indexing and retrieval speed, while other factors are assessed as part of overall system performance \cite{caspari_beyond_2024, kukreja_performance_2024}.

\section{Evaluation of Retriever}
\label{evaluationretriever}

A crucial aspect of evaluating a RAG system is assessing the performance of its retrieval component. In our survey, we identified 24 different papers that specifically address the evaluation (see Table \ref{tab:retrieverpaper}). In two studies, researchers adjust the retriever parameters and evaluate the generator's performance to analyze how changes in the retriever affect the overall system \cite{yu_knowledge-centric_2024, kukreja_performance_2024}, while the remaining ones focus on evaluation methods tailored to assess the retriever itself. The primary goal of retriever evaluation is to determine whether the retrieved chunks are relevant to the given query. We identified seven distinct methods for determining document relevance, followed by the application of metrics as listed in table \ref{tab:retrieverpaper} to calculate a score reflecting the quality of the retriever. Among these, Mean Reciprocal Rank (MRR) and Discounted Cumulative Gain (NDCG) are the only metrics that account for the order of the retrieved documents.
Another approach evaluates additional the fairness of the retriever by examining whether the retrieved documents fairly represent protected groups \cite{wu_does_2024}. 
\vspace{-1ex}
\begin{table}[htbp]
\centering
 \caption{List of metrics focusing on retriever evaluation}
\label{tab:retrieverpaper}
\begin{tabular}{|p{2.6cm}|p{3.3cm}|p{1,6cm}|}
\hline
\textbf{Context relevance}  & \textbf{Metric}  & \textbf{Ref.} \\ 
\hline
Labeled  & NDCG & \cite{moreira_enhancing_2024}  \\
& Accuracy, MRR & \cite{wu_does_2024}  \\
 & Recall, MRR, Mean Average Precision (MAP), NDCG & \cite{cheng_coral_2024}  \\
& Recall & \cite{pu_customized_2024}   \\
& Recall, MRR, NDCG &  \cite{lin_irsc_2024} \\
& MRR, MAP, Hit Rate &  \cite{tang_multihop-rag_2024, wang_feb4rag_2024}  \\
& Precision, F1-Score, Recall &  \cite{schimanski_climretrieve_2024} \\
 &Only Context Relevance &  \cite{ru_ragchecker_2024} \\
 \hline
Keyword Labeling & Recall, Accuracy & \cite{liu_cofe-rag_2024}  \\
& Recall, Effective Information Rate & \cite{zhu_rageval_2024}  \\
\hline
Human judges  & Accuracy &  \cite{afzal_towards_2024}\\
\hline
LCS  & Only Context Relevance &  \cite{hui_uda_2024} \\
\hline

 \hline
 LLM Judge  & MRR & \cite{rackauckas_evaluating_2024}\\
& Context Utilization & \cite{friel_ragbench_2024} \\
& Precision, Recall & \cite{ding_vera_2024}  \\
& Only Context Relevance & \cite{saad-falcon_ares_2024, xie_rag_2024}   \\
\hline
 LLM judge (Indirect) & Precision, Recall, MRR, MAP, NDCG, Hit Rate & \cite{salemi_evaluating_2024}\\
& Only Context Relevance & \cite{alinejad_evaluating_2024, es_ragas_2023, meyur_weqa_2024}\\
 \hline
No Context Relevance & Retrieval Time & \cite{kukreja_performance_2024}   \\
\hline
\end{tabular}%
\end{table}
\vspace{-1ex}

For evaluation, the \textbf{context relevance} metric is used to calculate most of the other metrics. This metric determines whether a context is relevant or not. In the following, the different methods to assess the relevance of a context are presented.

The simplest way to determine if a document is relevant to a question is by using a \textbf{labelled dataset}. In such datasets, each question is paired with a set of relevant contexts. A document is considered relevant if it matches the contexts labelled as relevant in the dataset \cite{cheng_coral_2024, pu_customized_2024}. However, these datasets often contain only a predefined set of documents, leading to errors if truly relevant documents are excluded and wrongly labeled as irrelevant \cite{liu_cofe-rag_2024, zhu_rageval_2024}.
To address this issue, some approaches label each context with a \textbf{keyword}, and each question is also associated with these keywords (see Section \ref{GenerationOfDataset}). If the retrieved context contains the keyword associated with the question, the context is deemed relevant \cite{liu_cofe-rag_2024, zhu_rageval_2024}.

In cases where no labelled dataset is available, \cite{afzal_towards_2024} relies on\textbf{ human judges} to determine whether the context is relevant. Additionally, in some instances, an extra human preference validation set is created to compare against the newly developed approach for evaluation purposes \cite{saad-falcon_ares_2024, xie_rag_2024}.
In \cite{hui_uda_2024}, \textbf{traditional retrieval evaluation strategies} are used. The ground truth evidence is available, and the Longest Common Subsequence (LCS) is calculated to measure the quality of the retriever.
One method of measuring context relevance is to employ an \textbf{LLM as judge} to calculate a score indicating whether the retrieved chunk is relevant. There are different methods to use the LLM to measure this.
One method involves using the \textbf{LLM directly} as a binary classifier to assess whether a document is relevant to the question, with only slight variations in the prompts used for different evaluations.\cite{saad-falcon_ares_2024}.
The eRAG framework \cite{salemi_evaluating_2024} evaluates chunk relevance using an \textbf{LLM indirectly} by generating a question from each chunk and assessing the correctness of the LLM's answer, labeling the context as relevant if accurate. Another approach evaluates relevance by comparing answers generated from retrieved documents to those derived from golden documents, which are essential for solving the question \cite{alinejad_evaluating_2024}. In RAGAS, relevance is measured by the proportion of sentences used from the provided context \cite{es_ragas_2023}.

\section{Evaluation of Generator}
\label{evalautionofgenerator}
This section outlines the evaluation of the generator based on a review of 56 relevant papers. These studies were grouped into four categories: \textbf{short answer evaluation}, focusing on methods for assessing multiple-choice or categorization tasks; \textbf{classical and embedding-based approaches}, representing traditional evaluation techniques predating large language models; \textbf{LLMs as evaluators}, where large language models perform evaluations; and \textbf{human evaluation}, involving assessments conducted by human judges.

\subsection{Short Answers Evaluation}
\label{evalautiongenerationofshort}
\vspace{-1ex}
\begin{table}[htbp]
\caption{List of metrics for Short Answer evaluation }
\label{tab:listofmetricsshortevalaution}
\begin{tabular}{|p{3cm}|p{5cm}|}
\hline
\textbf{Metrics}                                                           & \textbf{Papers}                                                                                                                                                                                                                                                                                           \\
\hline
F1-Micro Score                                                    & \cite{katranidis_faaf_2024}                                                                                                                                                                                                                                                    \\
 Precision                                                         & \cite{simon_methodology_2024,  rau_bergen_2024}                                                                                                                                                                                                                             \\
 Recall                                                            & \cite{simon_methodology_2024,  rau_bergen_2024, katranidis_faaf_2024, khaled_evaluating_2024}                                                                                                                                                                            \\
 Error Rejection/Detection/ Correction Rate & \cite{chen_benchmarking_2023}                                                                                                                                                                                                                                                  \\
 F1-Score                                                          & \cite{simon_methodology_2024, khaled_evaluating_2024, yu_reeval_2024, hui_uda_2024}                                                                                                                                                                                                      \\
 Accuracy                                                          & \cite{guinet_automated_2024, yang_crag_2024, xiong_benchmarking_2024, chen_benchmarking_2023, wu_does_2024, oberst_how_2023, yu_reeval_2024, tang_multihop-rag_2024}                                                                                             \\
 Error Rate                                                        & \cite{katranidis_faaf_2024}                                                                                                                                                                                                                                                    \\
\hline
\end{tabular}
\end{table}
\vspace{-2ex}
One approach to evaluating the RAG system is to simply determine whether the answer is correct. This can be automated when the answers are very short.
The correctness of an answer is measured using exact match metrics. This approach applies to tasks such as short answers \cite{chen_benchmarking_2023}, multiple-choice \cite{guinet_automated_2024} and binary \cite{simon_methodology_2024} or multi-class categorization \cite{khaled_evaluating_2024}. Evaluation datasets include the correct answers or categories, enabling automated detection of whether the RAG's response is correct. Based on this, the relevant metrics are calculated, as summarized in Table \ref{tab:listofmetricsshortevalaution}. This involves noisy datasets to calculate error metrics to test the model's robustness when handling noisy or erroneous input data \cite{chen_benchmarking_2023}.
This evaluation approach is commonly used in simpler RAGs, where answers can be easily assessed as correct or incorrect. However, for more complex RAGs, where responses are not as straightforward, additional methods are required for comprehensive evaluation.

\subsection{Human Evaluators}
\vspace{-2ex}
\begin{table}[htbp]
\caption{List of metrics for Human Evaluators}
\label{tab:listofmetricshumanevaluators}
\begin{tabular}{|p{1,6cm}|p{3,7cm}|p{2,3cm}|}
 \hline
 \textbf{Method}                                       & \textbf{Metrics}                                                           & \textbf{Papers}   \\

\hline
 Direct                                  & Correctness                                                       & \cite{lang_automatic_2024, sun_multimodal_2024, alghisi_should_2024, afzal_towards_2024}                                                                                                                                                                                 \\
   & Precision, Recall, Accuracy                                    & \cite{pipitone_legalbench-rag_2024, onan_benchmarking_2024}                                                                                                                                                                                                                  \\
   & Quality                                                           & \cite{lang_automatic_2024, afzal_towards_2024}                                                                                                                                                                                                                               \\
& Readability, Usefulness                                        & \cite{afzal_towards_2024}                                                                                                                                                                                                                                                      \\
&              Confusing Questions Detection                                                     & \cite{peng_rag-confusionqa_2024}                                                                                                                                                                                                                                               \\
   & Score                                                             & \cite{yang_crag_2024}                                                                                                          \\           \hline                                                                                                              
 Comprehensive                                  & Coverage, Consistency, Correctness, Clarity                   & \cite{wang_feb4rag_2024}                                                                                                                                                                                                                                                       \\
 \hline
\end{tabular}%
\end{table}
\vspace{-2ex}

In eleven approaches, human evaluators were employed to manually assess the responses of RAG systems. We identified two distinct methods for assessing the quality of responses: The first method involves direct evaluation, where human evaluators assess the quality of responses based on various metrics, as listed in Table \ref{tab:listofmetricshumanevaluators}, along with the corresponding papers. For example, one study tested the RAG system's ability to handle confusing questions by deliberately introducing errors in the context. Human evaluators then determined whether the RAG correctly identified or addressed these errors \cite{peng_rag-confusionqa_2024}. The second method is the comprehensive evaluation, introduced in Feb4Rag \cite{wang_feb4rag_2024}. In this approach, two answers are presented side by side, and human evaluators judge which answer is superior based on predefined metrics. Human evaluation was especially applied in domain-specific contexts. For example, in the Legal-Bench RAG, domain experts assessed the correctness of responses to legal questions. These evaluations were then used to calculate metrics \cite{pipitone_legalbench-rag_2024, onan_benchmarking_2024}. Additionally, eight approaches utilized human judgment as a benchmark to compare the performance of RAG systems against methods where an LLM acted as a judge \cite{leng_best_2023, wang_evaluating_2024, han_rag-qa_2024, afzal_towards_2024, ru_ragchecker_2024, rackauckas_evaluating_2024} or against classical evaluation methods \cite{yu_reeval_2024, afzal_towards_2024, lang_automatic_2024}.

\subsection{Classical and Embedding-Based Approaches}

\label{evalautionofgeneratorclassic}
\vspace{-2ex}
 \begin{table}[htbp]
 \caption{List of metrics for Embedding Based approaches}
 \label{tab:listofmetricsembedding}
\begin{tabular}{|p{1,5cm}|p{3cm}|p{3cm}|}
 \hline                                                                                                        \textbf{Method}     & \textbf{Metrics}          & \textbf{Papers}   
                        \\
\hline
Embedding                              & SAS                                                               & \cite{kukreja_performance_2024}                                         \\
& SBERT                                                             & \cite{yu_reeval_2024, afzal_towards_2024}                                                                                                                                                                                                                                    \\
& BERTScore                                                         & \cite{afzal_towards_2024, lyu_crud-rag_2024}                                                                                                                                                                                                                                 \\
\hline
 N-Gram                                 & BLEU                                                              & \cite{liu_cofe-rag_2024, cheng_coral_2024, pu_customized_2024, afzal_towards_2024, rackauckas_evaluating_2024}                                                                                                                                                         \\

 & ROGUE-n                                                           & \cite{wu_does_2024, kukreja_performance_2024, afzal_towards_2024, rau_bergen_2024}                                                                                                                                                                                       \\
  & ROGUE-L                                                           & \cite{liu_cofe-rag_2024, cheng_coral_2024, pu_customized_2024, wang_domainrag_2024, thakur_mirage-bench_2024, kukreja_performance_2024, afzal_towards_2024, rau_bergen_2024, rackauckas_evaluating_2024}                                                       \\
& Unigram Precision/Recall                                      & \cite{b_evaluating_2024}                                                                                                                                                                                                                                                       \\
\hline
Model                                   & Unieval                                                           & \cite{pu_customized_2024}                                                                                                                                                                                                                                                      \\
\hline
Token                                  & Similarity                                                        & \cite{li_benchmarking_2024, b_evaluating_2024}                                                                                                                                                                                                                               \\
\hline
\end{tabular}
\end{table}
\vspace{-2ex}

To automatically evaluate the performance of the generator in a RAG system, the generated response is compared to the ground truth answer. We identified four distinct methods for automatically evaluating the generator's performance.
One widely used approach is \textbf{embedding}-based evaluation, which assesses the semantic similarity between the generated answer and the reference answer. Within this category, we identified three primary techniques. The first is the Semantic Answer Similarity (SAS) score, which employs a trained cross-encoder architecture to evaluate the semantic alignment between generated and reference answers \cite{kukreja_performance_2024}. The second technique involves Sentence-BERT (SBERT), a specialized adaptation of the BERT model designed for comparing sentences and evaluating their semantic similarity \cite{devlin_bert_2019, reimers_sentence-bert_2019}. SBERT is particularly effective for tasks requiring textual alignment, making it suitable for comparing RAG-generated answers with reference answers \cite{yu_reeval_2024, afzal_towards_2024}. Lastly, BERTScore, another embedding-based model built upon the BERT architecture, compares token-level embeddings of sentences to measure similarity \cite{zhang_bertscore_2020, afzal_towards_2024, lyu_crud-rag_2024}.
In addition, \textbf{n-gram}-based metrics are commonly used. These metrics calculate scores automatically by analyzing sequences of words (n-grams) of a specified length. Table \ref{tab:listofmetricsembedding} lists the specific metrics used in this category.
The third method involves \textbf{model}-based evaluation, such as the UniEval score. UniEval is designed to assess natural language generation by providing a comprehensive evaluation score based on coherence, consistency, fluency, and relevance. In this approach, a model is trained using the ground truth as a reference for evaluation \cite{zhong_towards_2022, pu_customized_2024}.
Finally, the fourth method is \textbf{token-based} evaluation, which measures the quality of a generated response by calculating the ratio of overlapping tokens between the generated answer and the ground truth, using word segmentation tools \cite{li_benchmarking_2024}. Alternatively, this can be achieved using cosine similarity, which measures the cosine of the angle between two vectors \cite{b_evaluating_2024}. 
In four cases, these metrics are extended by using a large language model (LLM) as a judge \cite{liu_cofe-rag_2024, thakur_mirage-bench_2024, afzal_towards_2024, rau_bergen_2024} or a human evaluator for validation \cite{yu_reeval_2024, afzal_towards_2024}.

\subsection{LLM as a Judge}

\begin{table}[htbp]
\caption{List of metrics for LLM as a judge}
\label{tab:listofmetricsllmasjudge}
\begin{tabular}{|p{1,3cm}|p{2.7cm}|p{3.7cm}|}
 \hline
 \textbf{Method}                                       & \textbf{Metrics}                                                           & \textbf{Papers}   \\
 \hline
 EM+LLM                            & Correctness                                                       & \cite{yu_knowledge-centric_2024, yang_crag_2024}        \\
 & Error Explanation & \cite{chen_benchmarking_2023}
 \\
 \hline
Direct                                       & Faithfulness/Factual Consistency                                  & \cite{es_ragas_2023, saad-falcon_ares_2024, liu_cofe-rag_2024, wang_domainrag_2024, langchain_evaluation_nodate, huang_evaluation_2024, ding_vera_2024, ru_ragchecker_2024, kuo_rad-bench_2024, ming_faitheval_2024}                                       \\
& Truthfulness/Correctness                                          & \cite{rau_bergen_2024, liu_cofe-rag_2024, yang_crag_2024, xie_rag_2024, rackauckas_evaluating_2024, langchain_evaluation_nodate, krishna_fact_2024, sivasothy_ragprobe_2024, han_rag-qa_2024, kenneweg_retrieval_2024, hui_uda_2024, afzal_towards_2024} \\
 & Hallucination                                                     & \cite{yu_knowledge-centric_2024, yang_crag_2024, zhu_rageval_2024, ru_ragchecker_2024}                                                                                                                                                                                   \\
 & Relevance                                                         & \cite{es_ragas_2023, saad-falcon_ares_2024, liu_cofe-rag_2024, rackauckas_evaluating_2024, huang_evaluation_2024, zhu_rageval_2024, sivasothy_ragprobe_2024, kenneweg_retrieval_2024, afzal_towards_2024, ding_vera_2024}                                    \\
 & Redundancy                                                        & \cite{wang_domainrag_2024}                                                                                                                                                                                                                                                     \\
 & Noise Sensitivity                                                 & \cite{es_ragas_2023, ru_ragchecker_2024, ming_faitheval_2024}                                                                                                                                                                                                              \\
   & Completeness                                                       & \cite{rackauckas_evaluating_2024, zhu_rageval_2024, sivasothy_ragprobe_2024, han_rag-qa_2024}                                                                                                                                                                            \\
 & Precision                                                         & \cite{rau_bergen_2024}                                                                                                                                                                                                                                                         \\
 & Helpfulness                                                       & \cite{langchain_evaluation_nodate, han_rag-qa_2024}      \\                                                                                                                                                        & Missing                                                           & \cite{yu_knowledge-centric_2024, yang_crag_2024}                                                                                                                                                                                                                             \\
 & Deficiency                                                       & \cite{wang_domainrag_2024}                                                                                                                                                                                                                                                     \\
  & Coherence                                                          & \cite{kuo_rad-bench_2024}                                                                                                                                                                                                                                                      \\
 & Score                                                             & \cite{yu_knowledge-centric_2024, huang_evaluation_2024, yang_crag_2024, leng_best_2023, wang_evaluating_2024}                                                                                                                                                          \\
\hline
Indirect                                     & Precision, recall                                                 & \cite{lyu_crud-rag_2024}                                                                                                                                                                                                                                                       \\
               & Relevance                                                         & \cite{es_ragas_2023}                                                                                                                                                                                                                                                           \\
  & KPR                                                               & \cite{qi_long2rag_2024}                                                                                                                                                                                                                                                                        \\
    & Fact/Logic Consistency                                        & \cite{xu_face4rag_2024}                                                                                                                                                                                                                                                        \\
    & OPI                                                               & \cite{hu_intrinsic_2024}                                                                                                                                                                                                                                                       \\
\hline
Comparative                                  & Kendall´s tau                                                     & \cite{saad-falcon_ares_2024}                                                                                                                                                                                                                                                   \\
                        & RAGElo                                                            & \cite{arthur_zetaalphavectorragelo_2024, rackauckas_evaluating_2024, thakur_mirage-bench_2024}                                                                                                                                                                             \\
\hline
Own LLM                                      & Lynx                                                              & \cite{ravi_lynx_2024}                                                                                                                                                                                                                                                          \\
\hline
\end{tabular}
\vspace{-6ex}
\end{table}
\vspace{-1ex}

LLMs can serve as judges to evaluate the performance of RAG systems, and we identified 41 papers that employed LLMs for this purpose. Table \ref{tab:listofmetricsllmasjudge} lists all metrics and the corresponding papers used for evaluating the generator with an LLM. We observed a growing trend in using LLMs for automating evaluation. Initial studies have shown a positive correlation between human evaluation and LLM-based evaluation \cite{leng_best_2023, wang_evaluating_2024}, highlighting the potential of LLMs for assessing RAG systems. Based on our analysis, we identified five distinct methods for calculating these metrics.

The first method integrates the exact match (EM) metric, referred to as\textbf{ LLM + EM}. This approach leverages an LLM when the exact match fails, such as in cases where the generated answer is too long. In such instances, the LLM determines whether the answer is correct \cite{yu_knowledge-centric_2024, yang_crag_2024}. Additionally, the LLM can provide detailed explanations about the errors, offering a more nuanced understanding of the RAG system’s capabilities. For example, it can highlight the RAG’s ability to reflect information absent in the document or to identify factual inaccuracies \cite{chen_benchmarking_2023}.

The second approach involves \textbf{direct} measurement, where an LLM is prompted to calculate a specific metric directly. This method takes the generated answer, the question, the retrieved context, and the ground truth answer as input to compute the metric \cite{saad-falcon_ares_2024}. Here metrics like Faithfulness or Truthfulness were described in a specific prompt and the LLM outputs a score for these. Faithfulness describes for example the factual consistent by the retrieved context \cite{es_ragas_2023}. The direct measurement has been the most widely used method for evaluating the generator component.
In contrast, the \textbf{indirect} measurement method focuses on a more complex evaluation process. Unlike direct measurement, this approach does not rely on the direct use of the context, question, and ground truth in the prompt. Instead, it preprocesses the generated answer in a specific manner before calculating the metric. For instance, RAGQuestEval evaluates the ground truth reference by generating questions derived from it and then determining whether the generated answers can correctly address these questions \cite{lyu_crud-rag_2024}. Other methods in this category include counting key points in the generated answer \cite{qi_long2rag_2024} or generating questions based on the answer itself \cite{es_ragas_2023}.

For \textbf{comparative} evaluation, metrics such as Kendall’s tau are commonly employed to compare different RAG configurations \cite{saad-falcon_ares_2024}. A more comprehensive framework, RAGElo, was also introduced, combining individual judgments with a comparative scoring mechanism \cite{arthur_zetaalphavectorragelo_2024, rackauckas_evaluating_2024, thakur_mirage-bench_2024}. This framework allows an LLM to evaluate two answers to the same question along with their retrieved contexts, determining which answer is better. Based on these comparisons, an Elo score is calculated, serving as a comparative metric for evaluating the performance of RAG systems \cite{rackauckas_evaluating_2024, thakur_mirage-bench_2024}.
Finally, the fifth method leverages a specialized \textbf{own LLM} for evaluation. For example, Lynx, an open-source LLM trained specifically to detect hallucinations, has been used as a standalone evaluator to assess the accuracy and reliability of RAG systems \cite{ravi_lynx_2024}.
These diverse methods illustrate the versatility of LLMs as evaluators.

\section{Conclusion}
A key challenge in the evaluation workflow lies in selecting an appropriate dataset. This process must address several issues. First, the dataset must be designed to prevent the LLM from leveraging its base knowledge to answer questions independently \cite{kenneweg_retrieval_2024}. The questions should also be neither too simple nor too general; instead, they should be challenging and aligned with the RAG system's intended use case. Additionally, the dataset should contain labels for the context relevance if the retriever evaluation requires these.
Another challenge is that every part of the evalution can be automated with an LLM. Here the challange lies in ensuring confidence in the automation process, especially in tasks involving LLMs. Unresolved questions persist, such as whether the quality of evaluation is compromised when an LLM generates questions, answers them, and ultimately evaluates its own output. This raises concerns about potential biases when an LLM evaluates itself, and whether the evaluation remains reliable without human involvement. Determining where human expertise is still necessary in this workflow remains a critical area for further exploration. 
The rapid development of LLMs presents another significant challenge, as advancements in models could render previous evaluation results invalid. A new model might produce entirely different outcomes, raising the issue of how to adapt prior results and establish a standard evaluation framework that remains consistent and independent of specific LLM versions. Additionally, the lack of a standardized prompt for LLM-based evaluations complicates the ability to make results comparable across different studies.

In light of these evolving challenges, this study identifies and synthesizes best practices across key components of RAG system evaluation.
The indexing component is evaluated primarily based on its performance metrics and as part of the overall system performance \cite{caspari_beyond_2024, kukreja_performance_2024}.
The retriever's quality is assessed by the relevance of retrieved documents, often using labeled datasets \cite{salemi_evaluating_2024, moreira_enhancing_2024}. In the absence of such datasets in RAG systems, domain experts manually evaluated relevance \cite{afzal_towards_2024}, or LLMs were employed for efficient, automated assessment \cite{salemi_evaluating_2024}.
The generator component was evaluated through four methods: exact match detection for short answers \cite{katranidis_faaf_2024}, traditional metrics like BLEU or ROUGE for complex questions \cite{afzal_towards_2024}, human validation by experts for accuracy and relevance \cite{lang_automatic_2024}, and LLM-based evaluation using metrics like faithfulness \cite{es_ragas_2023}. The trend and best practice point toward using an LLM to automate this evaluation task.
This study also highlighted the crucial role of datasets in evaluating RAG systems. Three types of datasets were identified: existing datasets, enhanced datasets, and newly created datasets \cite{pipitone_legalbench-rag_2024, xu_face4rag_2024, es_ragas_2023}. The creation of completely new datasets was particularly significant for domain-specific RAG systems where no publicly available datasets existed \cite{wang_domainrag_2024}.
In general, the study demonstrated that almost every aspect of the evaluation process could potentially be automated using LLMs. Datasets can be generated automatically, while both answers and retrieved chunks can be evaluated by LLMs. However, human expertise remains essential for specific tasks, particularly in domain-specific RAG systems where nuanced judgment is required. 

Only six studies compared LLM judges with human judges, finding a positive correlation between their evaluations \cite{leng_best_2023, wang_evaluating_2024, han_rag-qa_2024, afzal_towards_2024, ru_ragchecker_2024, rackauckas_evaluating_2024}.
While promising, these findings underscore the ongoing need for human oversight. Future research should further investigate the reliability of LLM-based evaluation, as current evidence suggests that LLMs can be trusted to some extent, but their validity remains to be thoroughly established.

In conclusion, while significant progress has been made in RAG systems and their evaluation, the practical dos and don'ts remain largely unexplored, especially in real-world application domains. As businesses continue to invest in RAG implementations, the need for practical frameworks and actionable recommendations has become increasingly urgent. A notable gap in current evaluation approaches is the lack of consideration for RAG-specific requirements, such as ensuring systems stay updated with new knowledge or how retrievers effectively incorporate and maintain access to the most current information. Further research should address these gaps by offering clear guidance and developing practical tools to enhance the effectiveness and adaptability of RAG systems in dynamic, real-world environments.
\bibliographystyle{IEEEtran}
\bibliography{IEEEabrv,references}

\end{document}